\documentclass[11pt,letterpaper]{article}
\usepackage[margin=0.8in]{geometry}
\usepackage{ifpdf}
\ifpdf
    \usepackage[pdftex]{graphicx}
    \usepackage[update]{epstopdf}
\else
	\usepackage{graphicx}
\fi
\usepackage{hyperref}
\usepackage{wrapfig,bbm}
\usepackage{enumitem,color}
\usepackage{amsthm}
\newtheorem{theorem}{Theorem}
\newtheorem{lemma}{Lemma}
\newtheorem{corollary}{Corollary}

\newtheorem{prop}{\textbf{Proposition}}
 
\usepackage{array}
\usepackage{amsmath,amsthm}
\usepackage{amssymb}
\usepackage{amstext}
\usepackage{cite,setspace}

\newenvironment{IEEEproof}{\proof}{\endproof}

\begin{document}

\title{Caching and Delivery via Interference Elimination}
\author{Chao Tian and Jun Chen
\thanks{Chao Tian is with the Department of Electrical Engineering and Computer Science, The University of Tennessee, Knoxville, TN 37996, USA (email: chao.tian@utk.edu). Jun Chen is with the Department of Electrical and Computer Engineering, McMastr University, Hamilton, ON L8S 4L8, Canada (email: junchen@ece.mcmaster.ca). }}
\maketitle

\begin{abstract}
We propose a new caching scheme where linear combinations of the file segments are cached at the users, for the cases where the number of files is no greater than the number of users. When a user requests a certain file in the delivery phase, the other file segments in the cached linear combinations can be viewed as interferences. The proposed scheme combines rank metric codes and maximum distance separable codes to facilitate the decoding and elimination of these interferences, and also to simultaneously deliver useful contents to the intended users. The performance of the proposed scheme can be explicitly evaluated, and we show that the tradeoff points achieved by this scheme can strictly improve known tradeoff inner bounds in the literature; for certain special cases, the new tradeoff points can be shown to be optimal.    
\end{abstract}

\section{Introduction}

Caching is a natural data management strategy when communication has a bursty characteristic. During off-peak time, local cache can be filled with data that is anticipated to be useful later to reduce the delay when the communication resources become scarce during peak time. 

\begin{figure}[tb]
\centering
\includegraphics[width=10cm]{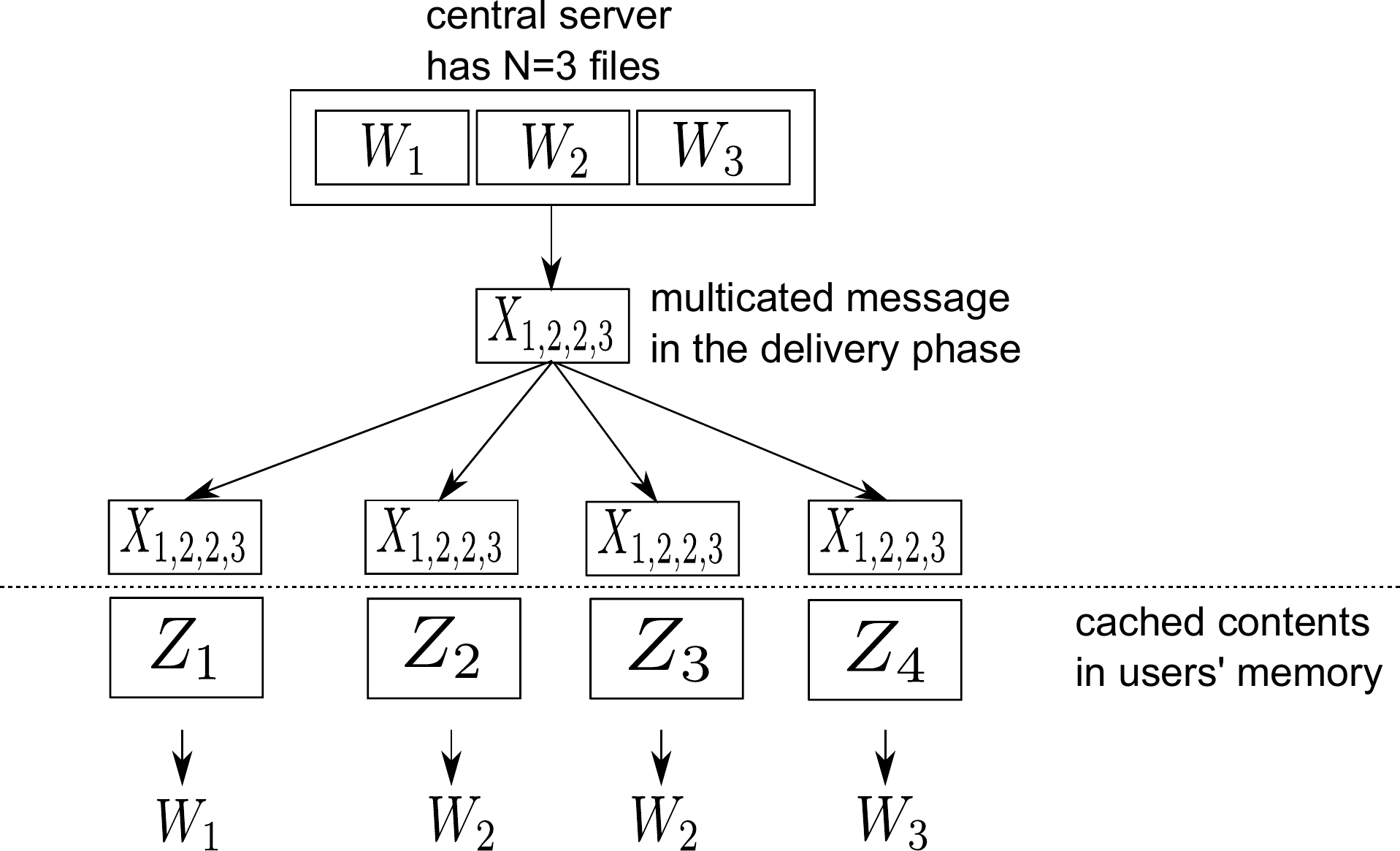}
\caption{An example caching system instance, where there are $N=3$ files and $K=4$ users. In this instance the users request files $(1,2,2,3)$, respectively, and thus the transmitted common information is written as $X_{1,2,3,4}$.\label{fig:system}}
\end{figure}

In a recent work \cite{MaddahAliNiesen:14}, Maddah-Ali and Niesen provided a formal information theoretic formulation for the caching problem. In this formulation, there are $N$ files, each of $F$ bits, and $K$ users. Each user has a local cache memory of capacity $M$ (measured in multiples of $F$). In the caching phase, the users can fill their caches with contents from the central server without the knowledge of the precise requests. In the delivery phase, each user will request one file from the central server, and the central server must multicast certain common (and possibly coded) information to all the users in order to accommodate these requests; an example case is given in Fig \ref{fig:system}. Since in the caching phase, the requests at the later phase are unknown, the cached contents must be strategically prepared at all the users. The goal is to minimize the amount of multicast information which has rate $R$ (also measured in multiples of $F$), under the constraint on cache memory $M$. It was shown in \cite{MaddahAliNiesen:14} that coding can be rather beneficial in this setting, while uncoded solutions suffer a significant loss. Subsequent works extended it also to decentralized caching placements \cite{MaddahAliNiesen:14Networking}, caching with nonuniform demands \cite{MaddahAliNiesen:14infocom}, online caching placements \cite{Pedarsani:14}, and hierarchical coded caching \cite{Nikhil:14}; the caching methods have also found their applications in device-to-device communication systems \cite{Ji:JSAC16}.

The scheme given in \cite{MaddahAliNiesen:14} utilizes uncoded caching and coded transmission. A close inspection of the performance of the scheme reveals that when $N\leq K$, many individual tradeoff points achieved by the scheme in \cite{MaddahAliNiesen:14} are not on the lower convex envelope, and thus an effective scheme is lacking for this case, particularly when the cache capacity is small. Though the scheme in \cite{MaddahAliNiesen:14} was shown to be within a constant factor of the optimum, the loss of efficiency can be relatively significant when either $N$ or $K$ is small. Particularly, for more sophisticated caching scenarios, usually either files or users need to be classified into smaller groups (see {\em e.g.} \cite{MaddahAliNiesen:14infocom}), and such loss of efficiency may be magnified. Recently Chen {\em et al.} \cite{Chen:14} extended a special scheme given in \cite{MaddahAliNiesen:14} for the case of $N=K=2$ to the case $N\leq K$, and showed that the tradeoff pair $\left(\frac{1}{K},\frac{N(K-1)}{K}\right)$ is achievable, which is in fact one of the optimal tradeoff points. 

In this work, we propose a new coded caching scheme when $N\leq K$ that caches linear combinations of the file segments. When files are not being requested by a user, their segments in the cached linear combinations can be considered as interferences by this user. Our scheme strategically eliminates these interferences by utilizing a combination of rank metric codes and maximum distance separable codes; the transmission also simultaneously serves the role of content delivery to other users. We show that the proposed scheme provides new tradeoff points outside the known achievable tradeoff inner bound in the literature. In fact, in certain cases, it can achieve points on the optimal tradeoff function. In contrast to previous schemes in the literature, the proposed codes are not binary, but in larger finite fields. One disadvantage of utilizing rank metric codes is the large field size that the codes require, however we show that by directly considering the underlying rank constraints and utilizing generic linear codes, a smaller field size is sufficient for such codes to exist.

In the rest of the paper, we shall first give the main theorem in Section \ref{sec:maintheorem}, then introduce some preliminaries in Section \ref{sec:pre}. Before presenting the new codes, we provide three examples to illustrate the design principles in Section \ref{sec:examples}. The coding scheme, the corresponding proofs of correctness and analysis are given in Section \ref{sec:scheme} and Section \ref{sec:proof}, respectively. We conclude the paper in Section \ref{sec:conclusion}, and relegate some more technical proofs to the appendix.

\section{Main Theorem}
\label{sec:maintheorem}
The main result of this paper is summarized below,  where $\mathbb{N}$ used to denote the set of natural numbers.
\begin{theorem}
For $N\in \mathbb{N}$ files and $K\in \mathbb{N}$ users each with a cache of size $M$, where $N\leq K$, \label{theorem:main} the following $(M,R)$ pair is achievable
\begin{align}
\left(\frac{t[(N-1)t+K-N]}{K(K-1)},\frac{N(K-t)}{K}\right), \qquad t=0,1,\ldots,K.
\end{align}
\end{theorem}

With $t=0$ the tradeoff point degenerates to the trivial one $(M,R)=(0,N)$, {\em i.e.,} no cache; when $t=1$, it gives the same tradeoff pair as given in \cite{Chen:14}; when $t=K$, we obtain another trivial point of $(M,R)=(N,0)$, {\em i.e.,} no delivery transmission.
Together with the result in \cite{MaddahAliNiesen:14}, which is replicated in the next section (see Theorem \ref{theorem:AliNiesen}), we have the following corollary.

\begin{figure}[tb]
\centering
\includegraphics[width=10cm]{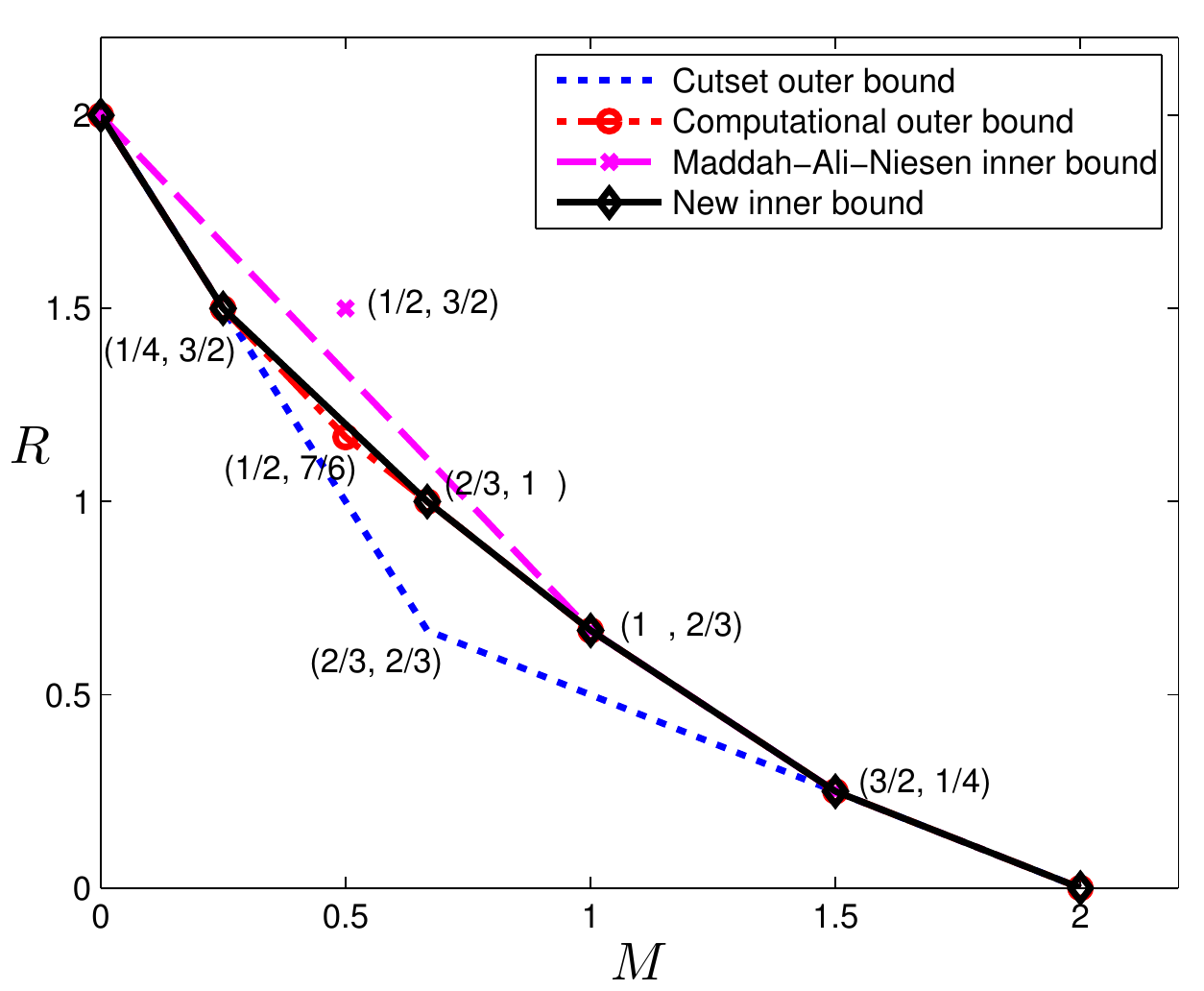}
\caption{Illustration of the tradeoff inner bounds and the outer bounds for $(N,K)=(2,4)$. The new inner bound matches the computation-based outer bound when $M\in [0,1/4]\cup[2/3,2]$. The pair $(1/2,3/2)$ can be achieved by the scheme given in \cite{MaddahAliNiesen:14}, but it is not on the convex envelope of the known inner bound.\label{fig:case2_4}}
\end{figure}

\begin{corollary}
For any $N\in \mathbb{N}$ and  $K\in \mathbb{N}$ where $N\leq K$, the lower convex envelope of the points in Theorem \ref{theorem:main} and those in Theorem \ref{theorem:AliNiesen} for $0\leq M\leq N$ is achievable. \label{corollary:combined}
\end{corollary}

The new tradeoff inner bound is illustrated for the case $(N,K)=(2,4)$ in Fig. \ref{fig:case2_4}. It can be seen that the scheme strictly improves upon the inner bound given in \cite{MaddahAliNiesen:14}. For reference, the cut-set based outer bound \cite{MaddahAliNiesen:14} is also shown, together with a computation-based outer bound established in a separate work (see \cite{TianWebpage}) using a method developed in \cite{Tian:JSAC13}. The new scheme gives the left three corner points on the solid black line (labeled with diamonds). The first two are previously known, being the trivial case with no cache, and the point given in \cite{Chen:14}, respectively. The third point is previously unknown to be achievable, and it is explained in detail in Section \ref{sec:examples}. Here all three points given by the new code are in fact on the optimal tradeoff function. 

In the proposed scheme, for demands where not all files are requested, the scheme can be viewed as degenerate cases of the scheme for certain enhanced demands, where all files are being requested. Although the scheme for such demands can be viewed as degenerate, this does not imply the tradeoff points achieved by the proposed scheme is only effective when $R\geq N-1$, for which non-trivial codes are required only for the demands that all files are requested. An example is given for the case of $(N,K)=(4,20)$ to illustrate the different tradeoff points achieved by the proposed scheme and those achieved by the scheme given in  \cite{MaddahAliNiesen:14}. The lower convex hull specified in Corollary \ref{corollary:combined} consists of three regimes: the low memory regime where the proposed scheme dominates, a transition regime (red solid line) which is achieved by space sharing between the proposed scheme and the scheme in \cite{MaddahAliNiesen:14}, and a high memory regime where the scheme in \cite{MaddahAliNiesen:14} dominates. The point $(M,R)=(259/380, 13/5)$ in on the lower convex hull, and it can be seen that the transmission rate is less than $N-1=3$ here.

\begin{figure}[tb]
\centering
\includegraphics[width=10cm]{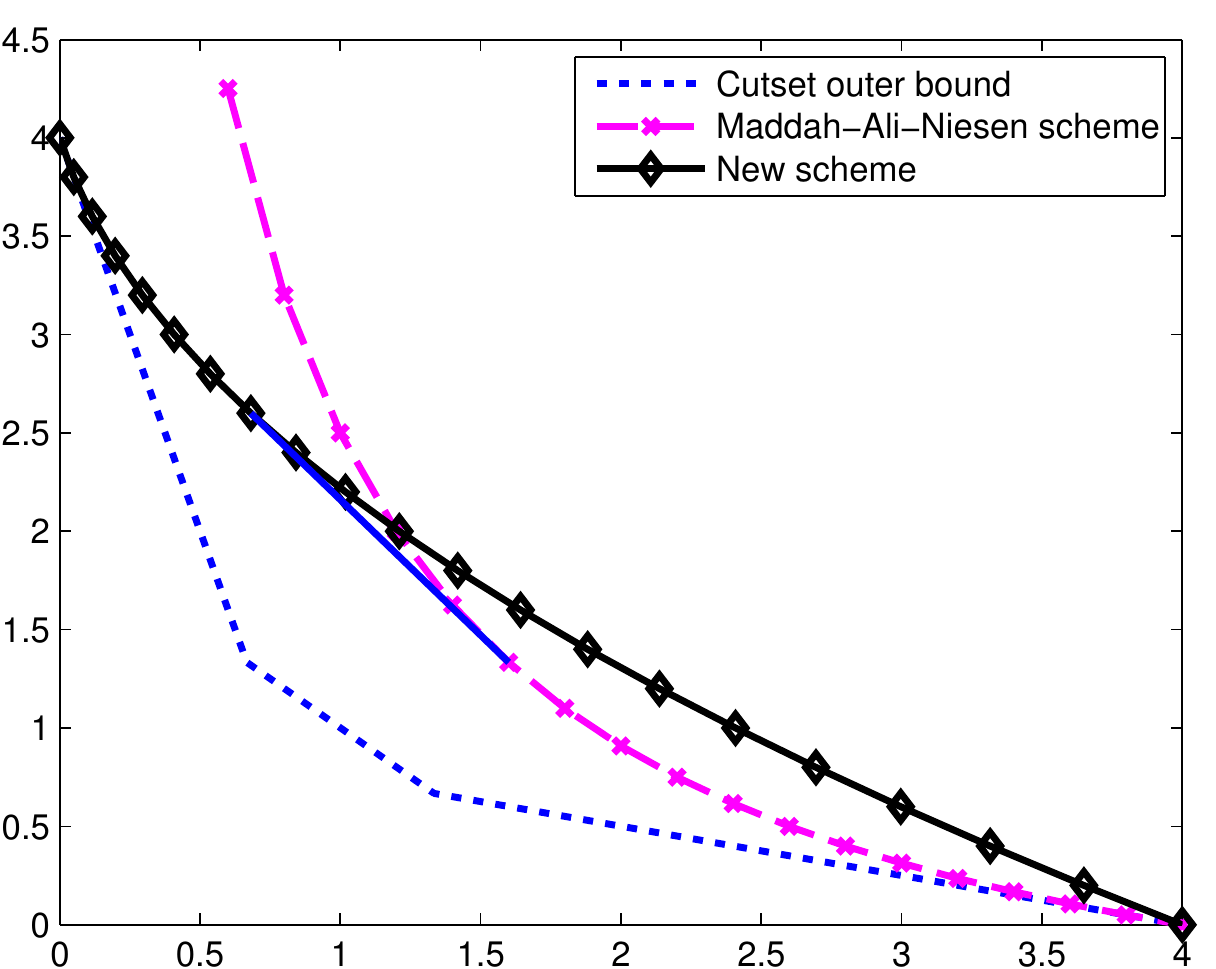}
\caption{Illustration of the achievable tradeoffs for $(N,K)=(4,20)$.\label{fig:case4_20}}
\end{figure}

\section{Preliminaries}
\label{sec:pre}

In this section we review some existing results on the caching problem, and then provide some necessary background information on maximum distance separable (MDS) codes and rank metric codes.  

\subsection{Known Caching Schemes and Achievable Tradeoffs}
\begin{theorem}[Maddah-Ali and Niesen] For $N\in \mathbb{N}$ files and $K\in \mathbb{N}$ users each with a cache of size $M\in \{0,N/K,2N/K,\ldots,N\}$, \label{theorem:AliNiesen}
\begin{align}
R=K(1-M/N)\cdot \min\left\{\frac{1}{1+KM/N},\frac{N}{K}\right\}
\end{align}
is achievable. For general $0\leq M\leq N$, the lower convex envelope of these $(M,R)$ points is achievable. 
\end{theorem}

The first term in the minimization is achieved by the scheme of uncoded caching together with coded transmission \cite{MaddahAliNiesen:14}, while the latter term is by simple uncoded caching and uncoded transmission. Though this theorem is indeed correct, it can be slightly misleading since it may give the impression that the simple uncoded caching and uncoded transmission scheme can be effective in certain regime when $N<K$. A close examination reveals that this trivial scheme only provides one operating point of $(N,0)$ in the convex hull when $N\leq K$, as illustrated in Fig. \ref{fig:case2_4}. Thus a good caching strategy for the low memory case is still lacking. 

As mentioned early, in a recent work \cite{Chen:14}, Chen {\em et al.} extended a special scheme for the case $N=K=2$ discussed in  \cite{MaddahAliNiesen:14} to the general case $N\leq K$, and showed that the tradeoff pair $\left(\frac{1}{K},\frac{N(K-1)}{K}\right)$ is achievable. It should be noted that the scheme given in \cite{MaddahAliNiesen:14} uses uncoded caching with coded transmission, while the scheme in \cite{Chen:14} uses coded caching and coded transmission. Both schemes use only binary coding, in contrast to the codes we propose in this work.

\subsection{Maximum Distance Separable Codes}
A linear code of length $n$ and dimension $k$ is called an $(n,k)$ code. The Singleton bound (see {\em e.g.}, \cite{Wicker:book}) is a well known upper bound on the minimum distance for any $(n,k)$ code, given as
\begin{align}
d_{\min}\leq n-k+1.
\end{align}

An $(n,k)$ code that satisfies the Singleton bound with equality is called a maximum distance separable (MDS) code. A key property of an MDS code is that it can correct any $(n-k)$ or fewer erasures \cite{Wicker:book}. For any $(n,k)$ pairs where $n\geq k$, MDS codes exist in any finite field $\mathbb{F}_q$ when $q\geq n$.

\subsection{Linearized Polynomial and Rank Metric Codes}
\label{subsec:pre}
In order to handle the competing coding requirements in the caching problem, we use rank metric codes based on linearized polynomials (see \cite{Gab85}), for which the following lemma is particularly relevant; see, {\em e.g.,} \cite{LidNie}. 

\begin{lemma} \label{lem:useful}
A linearized polynomial in finite field $\mathbb{F}_{q^m}$
\begin{align}
f(x)=\sum_{i=1}^{P}v_i x^{q^{i-1}}, \  v_i \in \mathbb{F}_{q^m}
\label{eqn:linearized}
\end{align}
can be uniquely identified from evaluations at any $P$ points $x=\theta_i\in \mathbb{F}_{q^m}$, $i=1,2,\ldots,P$, that are linearly independent 
over $\mathbb{F}_q$. 
\end{lemma}

Another relevant property of linearized polynomials is that they satisfy the following condition 
\begin{align}
f(ax + by)  =  af(x) + bf(y), \ a, b \in \mathbb{F}_q, \ x, y \in \mathbb{F}_{q^m},
\end{align}
which is the reason that they are called \lq\lq{}linearized\rq\rq{}. This property implies the following lemma.

\begin{lemma}
\label{lemma:fullrank}
Let $f(x)$ be a linearized polynomial in $\mathbb{F}_{q^m}$ as given in (\ref{eqn:linearized}), and let $\theta_i\in \mathbb{F}_{q^m}$, $i=1,2,\ldots,{P_o}$, be linearly independent over $\mathbb{F}_q$. Let $G$ be a $P_o\times P$ full rank (rank $P$) matrix with entries in $\mathbb{F}_q$, then $f(x)$ can be uniquely identified from 
\begin{align}
[f(\theta_1),f(\theta_2),\ldots,f(\theta_{P_o})]\cdot G. 
\end{align} 
\end{lemma}
\begin{IEEEproof}
We slightly abuse the notation by allowing the function $f(x)$ to take vector input in $\mathbb{F}^{P_0}_{q^m}$, and define the output as the vector obtained by concatenating the output of $f(x)$ on each input component. Then by the linearized property, 
\begin{align*}
[f(\theta_1),f(\theta_2),\ldots,f(\theta_{P_o})]\cdot G
=&[f(\theta_1,\theta_2,\ldots,\theta_{P_o})]\cdot G\\
=&f[(\theta_1,\theta_2,\ldots,\theta_{P_o})\cdot G].
\end{align*} 
Recall when each $\theta_i$ is viewed as a vector in $\mathbb{F}_{q}$, the $(\theta_1,\theta_2,\ldots,\theta_{P_o})$ vectors are linearly independent. Since $G$ has rank $P$,  $(\theta_1,\theta_2,\ldots,\theta_{P_o})\cdot G$ has rank $P$ in $\mathbb{F}_q$, {\em i.e.}, we have $P$ evaluations of $f(x)$ at $P$ linearly independent values, and thus by Lemma \ref{lem:useful}, $f(x)$ can be uniquely identified. 
\end{IEEEproof}

With a fixed set of $\theta_i\in \mathbb{F}_{q^m}$, $i=1,2,\ldots,{P_o}$, which are linear independent, 
we can view $(v_1,v_2,\ldots,v_P)$ as information symbols to be encoded, and the evaluations $[f(\theta_1),f(\theta_2),\ldots,f(\theta_{P_o})$] as the coded symbols. This is a $(P_o,P)$ MDS code in terms of rank metric. More importantly, the above lemma says any full rank (rank $P$) $\mathbb{F}_q$ linear combinations of the coded  symbols are sufficient to decode all the information symbols. This linear-transform-invariant property had been utilized previously in other coding problems such as network coding with errors and erasures \cite{Koetter:08}, locally repairable codes with regeneration \cite{Silberstein:15}, and layered regenerating codes \cite{Tian:15}.

The codes thus obtained are not systematic, but they can be converted to systematic codes by viewing the information symbols $(w_1,w_2,\ldots,w_P)$ as the first $P$ evaluations $[f(\theta_1),f(\theta_2),\ldots,f(\theta_{P})]$, which can be used to find the coefficients of the linearized polynomial $(v_1,v_2,\ldots,v_P)$, and then the additional parity symbols can be generated by evaluating this linearized polynomial at the remaining points $(\theta_{P+1},\ldots,\theta_{P_o})$. Systematic rank-metric codes are instrumental in our construction.

\section{Three Examples}
\label{sec:examples}
In this section, we provide three examples to illustrate the caching and transmission mechanism and discuss several critical observations. These observations provide important  intuitions, which are used to design the caching and transmission strategy for the general case. 

\begin{table*}[tb!]
\begin{center}
\caption{Caching content for $(N,K)=(2,4)$\label{tab:newcorner1}}
\begin{tabular}{|c || c | c | c | c|}
\hline
User 1 &$A_1+B_1$ & $A_2+B_2$ & $A_3+B_3$ & $A_1+A_2+A_3+2(B_1+B_2+B_3)$\\\hline
User 2 &$A_1+B_1$ & $A_4+B_4$ & $A_5+B_5$ & $A_1+A_4+A_5+2(B_1+B_4+B_5)$\\\hline
User 3 &$A_2+B_2$ & $A_4+B_4$ & $A_6+B_6$ & $A_2+A_4+A_6+2(B_2+B_4+B_6)$\\\hline
User 4 &$A_3+B_3$ & $A_5+B_5$ & $A_6+B_6$ & $A_3+A_5+A_6+2(B_3+B_5+B_6)$\\
\hline
\end{tabular}
\end{center}
\end{table*}

\subsection{A Code for $(N,K)=(2,4)$: The Rank Counting Perspective}

In this example, the two files are denoted as $A$ and $B$, each of which is partitioned into $6$ segments of equal size, denoted as $A_i$ and $B_i$, respectively, $i=1,2,\ldots,6$. The contents in the cache of each user are given in Table \ref{tab:newcorner1}. By the symmetry of the cached contents, we only need to consider the demand $(A,A,A,B)$, {\em i.e.}, the first three users requesting $A$ and user $4$ requesting $B$, and the demand $(A,A,B,B)$, {\em i.e.,} the first two users requesting $A$ and the other two requesting $B$. Assume the file segments are in $\mathbb{F}_{5}$, which is the field we operate. This code we present next can achieve $(M,R)=(\frac{2}{3},1)$ which is strictly outside the known achievable tradeoff, as illustrated in Fig. \ref{fig:case2_4}.

\begin{itemize}
\item For  the demands $(A,A,A,B)$, the transmission is as follows,
\begin{align*}
\text{Step $1$:  }&B_1,B_2,B_4;\\
\text{Step $2$: }&A_3+2A_5+3A_6,A_3+3A_5+4A_6;\\
\text{Step $3$:  }&A_1+A_2+A_4.
\end{align*}
After step $1$, user 1 can recover $(A_1,A_2)$; furthermore, he has $(A_3+B_3,A_3+2B_3)$ by eliminating known symbols $(A_1,A_2,B_1,B_2)$, from which $A_3$ can be recovered. After step $2$, he can obtain $(2A_5+3A_6,3A_5+4A_6)$ to recover $(A_5,A_6)$. Using the transmission in step $3$, he can obtain $A_4$ since he has $(A_1,A_2)$. User 2 and user 3 can use a similar strategy to reconstruct all file segments in $A$. User 4 only needs $B_3,B_5,B_6$ after step $1$, which he already has in his cache, however they are contaminated by file segments from $A$. Nevertheless, he knows $A_3+A_5+A_6$ by recognizing 
\begin{align}
&(A_3+A_5+A_6)=2\sum_{i=3,5,6}(A_i+B_i)\nonumber\\
&\qquad\quad-[A_3+A_5+A_6+2(B_3+B_5+B_6)].
\end{align}
Together with the transmission in step $2$, user 4 has three linearly independent combinations of $(A_3,A_5,A_6)$. After recovering them, he can remove these interferences from the cached content for $(B_3,B_5,B_6)$.

\item 
For the demand $(A,A,B,B)$, we can send 
\begin{align*}
\text{Step $1$:  }& B_1,A_6;\\
\text{Step $2$: }&A_2+2A_4,A_3+2A_5,B_2+2B_3,B_4+2B_5.
\end{align*}
User 1 has $A_1,B_1,A_6$ after step $1$, and he can also form
\begin{align*}
B_2+B_3=&[A_2+A_3+2(B_2+B_3)]\\
&\qquad\qquad -(A_2+B_2) -(A_3+B_3),
\end{align*}
and together with $B_2+2B_3$ in the transmission of step $2$, he can recover $(B_2,B_3)$, and thus $A_2,A_3$. He still needs $(A_4,A_5)$, which can be recovered straightforwardly from the transmission $(A_2+2A_4,A_3+2A_5)$ since he already has $(A_2,A_3)$. Other users can use a similar strategy to decode their requested files.
\end{itemize}

This example may seem rather complicated and arbitrary at first sight, however, we can make a few observations which should clarify the purpose of each transmission. 

The placement of the file segments has certain similarity to the scheme in \cite{MaddahAliNiesen:14}. Each file is partitioned into segments, and each segments are given to multiple users, however, they are stored only as linear combinations with segments from other files. The first several (3 in this example) symbols can be viewed as {\em semi-systematic}, as they are simple summations of the corresponding file segments, while the last symbol is a local parity symbol. However, it is not necessary to classify the cached contents at a user into these two categories, but we choose to present the example this way to facilitate presentation. In the next two examples and the general construction, we present the code in a more general manner.

Step $1$ is uncoded which provides certain segments to users that request it, but at the same time helps to eliminates some interferences at other users. A segment from a file is transmitted uncoded only when it is not present at any users\footnote{In the proposed scheme, a file segment is present in a user's cache only as a component in some linear combinations, however we shall simply refer to it as ``present'' at the user. } that are requesting this file. Step $2$ is coded transmission, and it also serves the dual role of interference elimination and content delivery. In this step, we only transmit linear combinations of segments, each of which is formed by linearly combining segments from a single file; in fact, each such combination is formed with symbols present at a single user that is not requesting this file. For example, for the case $(A,A,A,B)$, the transmission $A_3+2A_5+3A_6$ has symbols in the cache of user 4, but user 4 is not requesting file $A$. The coefficients of the linear combinations in caching and transmission need to be chosen carefully to guarantee certain full rank property; {\em cf.} again, the transmissions by user 4 for the case $(A,A,A,B)$ in the example. 

The most important observation is the following alternative view of the transmission and decoding process. Take for instance the case with demand $(A,A,A,B)$: user 4 receives symbols $(A_3+2A_5+3A_6,A_3+3A_5+4A_6)$, together with $4$ cached symbols, all of which are linear combinations of basis $(A_3,A_5,A_6,B_3,B_5,B_6)$. If these linear combinations are linearly independent, then all these symbols can be solved. A close inspection reveals they are indeed linearly independent, and in fact the decoding process at any given user can be understood this way. The precise linear combination coefficients are not important, however, the linear independence (or the coding matrix being full rank) directly leads to the resolution of all interferences. 
For this reason, in the next example we do not explicitly specify the linear combination coefficients, but only the basis of the subspace and the dimension. For this purpose, we introduce the linear subspace notation of 
\begin{align}
\mathcal{L}[\text{subset of files}; \text{index subset}; \text{dimension}],
\end{align}
which means a subspace of the given dimension with the basis being the segments from the given files with the given subscript indices.
For example, the subspace spanned by $(A_3+2A_5+3A_6,A_3+3A_5+4A_6)$ shall be written as $\mathcal{L}[A;\{3,5,6\};2]$, which means a dimension 2 linear subspace in the subspace with basis $(A_3,A_5,A_6)$. Further notice that if the dimension is chosen to be the same as the dimension of the subspace, it is equivalent to an uncoded transmission of this basis. We shall assume in the next example all necessary full rank properties can be satisfied by properly choosing the coefficients, and in the general scheme, we show that one particular choice of such coefficients based on linearized polynomials indeed exists.

\subsection{A Code for $(N,K)=(3,6)$: Efficient Interference Elimination}

Given the observations above, we shall from here on adopt the indexing method in \cite{MaddahAliNiesen:14}, and enumerate the file segments by the subset of users they are present at. For example when $(N,K)=(3,6)$, file $A$ has segments $A_{1,2,3},A_{1,2,4},$ etc., and $A_{1,2,3}$ is present at users $1,2$, and $3$ in some linear combinations; {\em i.e.,} we choose to place any file segment at $t=3$ nodes as a component of some linear combinations. In this example, we reserve the letter $\mathcal{S}$ to enumerate some subset $\mathcal{S}\subseteq \{1,2,...,6\}$ and $|\mathcal{S}|=3$, where $|\cdot|$ is used to denote the cardinality of a set. 
For the case of $K=6$, the $k$-th user caches the following the linear combinations of files $(A,B,C)$:
\begin{align*} 
\mathcal{L}[A,B,C;\{\mathcal{S}:k\in \mathcal{S}\};18],\quad k=1,2,\ldots,6, 
\end{align*}
where the dimension $18$ is chosen because the memory usage at this point is $9/10$ as in Theorem \ref{theorem:main}, and each file is partitioned into ${6 \choose 3}=20$ segments, which implies that each node should cache $18$ symbols.

We shall not discuss all the cases of file demands for this example because it is rather lengthy, but will consider one case, since it brings out a very important ingredient in our transmission strategy. 

Let us consider the case when the users request $(A,A,A,B,B,C)$. The transmissions in step $1$ are uncoded transmissions similarly as in the previous case, however let us focus our attention on users $4,5,6$ which are not requesting $A$, in the subsequent steps. After the transmissions in step 1, these users still have the file segments in Table \ref{tab:specialcase} as {\em interferences}, which need to be eliminated. 
Though we can transmit linear combinations of the basis
\begin{align}
A_{1,4,5},A_{2,4,5},A_{3,4,5},A_{1,4,6},A_{2,4,6},A_{3,4,6},
\end{align}
directly to eliminate this interferences at user 4, this strategy is not very efficient. Observe the following: the basis $(A_{1,4,5},A_{2,4,5},A_{3,4,5})$, which are labeled red in the table, are present in both user 4 and user 5; the basis $(A_{1,4,6},A_{2,4,6},A_{3,4,6})$, which are labeled blue, are at both user 4 and user 6; $(A_{1,5,6},A_{2,5,6},A_{3,5,6})$ are at user 5 and user 6. 
We can thus alternatively transmit 
\begin{align*}
\mathcal{L}[A;\{\{1,4,5\},\{2,4,5\},\{3,4,5\}\};2],\\
\mathcal{L}[A;\{\{1,4,6\},\{2,4,6\},\{3,4,6\}\};2],\\
\mathcal{L}[A;\{\{1,5,6\},\{2,5,6\},\{3,5,6\}\};2].
\end{align*}
Each of these subspaces provides $2$ dimensional reduction of the interferences at 2 users simultaneously. This results in a total of dimension 4 interference reduction at each user with transmission of 6 symbols, which is difficult to accomplish without taking advantage of these subspace intersections.

\begin{table}[tb!]
\begin{center}
\setlength{\tabcolsep}{5pt}
\caption{Interference pattern from file $A$ for $(N,K)=(3,6)$\label{tab:specialcase}}
\begin{tabular}{|c || c | c | c | c | c | c|}
\hline
User 4 &\textcolor{red}{$A_{1,4,5}$} & \textcolor{red}{$A_{2,4,5}$} & \textcolor{red}{$A_{3,4,5}$} & \textcolor{blue}{$A_{1,4,6}$} & \textcolor{blue}{$A_{2,4,6}$} & \textcolor{blue}{$A_{3,4,6}$} \\\hline
User 5 &\textcolor{red}{$A_{1,4,5}$} & \textcolor{red}{$A_{2,4,5}$} & \textcolor{red}{$A_{3,4,5}$} & $A_{1,5,6}$ & $A_{2,5,6}$ & $A_{3,5,6}$\\\hline
User 6 &\textcolor{blue}{$A_{1,4,6}$} & \textcolor{blue}{$A_{2,4,6}$} & \textcolor{blue}{$A_{3,4,6}$} & $A_{1,5,6}$ & $A_{2,5,6}$ & $A_{3,5,6}$\\\hline
\end{tabular}
\end{center}
\end{table}

\subsection{A Code for $(N,K)=(3,4)$: Degenerate File Requests}

In this example, there are three files $(A,B,C)$, and we choose the parameter $t=2$, {\em i.e.,} each file is partitioned into $6$ segments and each segment is placed at two nodes. We wish to show that the tradeoff pair $(\frac{5}{6},\frac{3}{2})$ is achievable by extending the code given in the previous examples, though this tradeoff point is actually worse than known results in the literature. Note that since $R\leq 2$, the types of demands where only two files are requested cannot be satisfied by simply transmitting these files directly. As it turns out, these cases can be considered as degenerate from the cases when all files are being requested by the users.

\begin{table*}[htb!]
\begin{center}
\caption{Caching content for the example $(N,K)=(3,4)$\label{tab:newcode33}}
\begin{tabular}{|c || c | }
\hline
User 1 &$\mathcal{L}[A,B,C;\{\{1,2\},\{1,3\},\{1,4\}\};5]$\\\hline
User 2 &$\mathcal{L}[A,B,C;\{\{1,2\},\{2,3\},\{2,4\}\};5]$\\\hline
User 3 &$\mathcal{L}[A,B,C;\{\{1,3\},\{2,3\},\{3,4\}\};5]$\\\hline
User 4 &$\mathcal{L}[A,B,C;\{\{1,4\},\{2,4\},\{3,4\}\};5]$\\\hline
\end{tabular}
\end{center}
\end{table*}

The three nodes cache the contents as shown in Table. \ref{tab:newcode33}. 
Only the following three types of requests need to be considered due to symmetry:
\begin{itemize}
\item For  the case $(A,A,B,C)$, the transmissions are as follows:
\begin{align*}
\text{Step $1$:  }&A_{3,4},B_{1,2},B_{1,4},B_{2,4},C_{1,2},C_{1,3},C_{2,3};\\
\text{Step $2$:  }&\mathcal{L}[A;\{\{1,3\},\{2,3\}\};1],\mathcal{L}[A;\{\{1,4\},\{2,4\}\};1].
\end{align*}
\item For  the case $(A,A,B,B)$, the transmissions are as follows:
\begin{align*}
\text{Step $1$:  }&A_{3,4},B_{1,2},B_{1,4},B_{2,4};\\
\text{Step $2$:  }&\mathcal{L}[A;\{\{1,3\},\{2,3\}\};1],\mathcal{L}[A;\{\{1,4\},\{2,4\}\};1];\\
\text{Step $4$: }&B_{1,3},B_{2,3},C_{1,2}.
\end{align*}
\item For the case $(A,A,A,C)$, the transmissions are as follows:
\begin{align*}
\text{Step $1$:  }&A_{3,4},C_{1,2},C_{1,3},C_{2,3};\\
\text{Step $2$:  }&\mathcal{L}[A;\{\{1,3\},\{2,3\}\};1],\mathcal{L}[A;\{\{1,4\},\{2,4\}\};1];\\
\text{Step $4$: }&A_{1,2},A_{1,4},B_{2,4}.
\end{align*}
\end{itemize}

It can be verified that these transmissions indeed fulfill all the demands by counting the rank reduction for the purpose of interference elimination, as discussed in the first example.  Next let us make a few more observations in this solution.

The transmission for the first case follows the  strategy we have identified in the first example, but the other two cases require additional attention. For those two cases, the first two steps are still in line with our previous example for $(N,K)=(2,4)$, but there is an additional Step 4, where uncoded transmissions are used. In fact, the transmissions in the first two steps for the latter two cases are precisely those in the first two steps for the first case, except that the transmissions involving files not being requested are omitted. In the transmissions of Step 4, instead of transmitting the segments of the file not being requested, the corresponding file segments from another file are transmitted, with a few exceptions when those substituted segments have already been transmitted; if this occurs, the corresponding segments from the file not being requested are in fact transmitted. 

We can view the transmissions in the latter two cases as a variation from that in the first case. Let us focus on the case $(A,A,B,B)$: the only difference from the case $(A,A,B,C)$ is that user 4 is requesting file $B$ instead of $C$. A closer examination of the case $(A,A,B,C)$ reveals that all transmissions involving file $C$ are uncoded. Now to build the transmissions for the case $(A,A,B,B)$ from the transmissions for the case $(A,A,B,C)$, we replace these uncoded transmissions with the matching transmissions of segments of file $B$, however, only when there is no redundancy in such transmissions. For example, the last symbol to be transmitted should have been $B_{1,2}$ with such a straightforward substitution, but since we have already transmitted $B_{1,2}$, retransmitting it is unnecessary  and wasteful; instead the file segment $C_{1,2}$ is transmitted. In this case although no user is requesting file $C$, the last transmission does not cause any essential loss. In summary, a case when only a subset of files are requested can be viewed as degenerate, for which the transmission strategy can be deduced from some other case when all files are requested.

\section{The General Coding Scheme}
\label{sec:scheme}

Before presenting the general coding scheme, we first clarify the notation that will be used in the sequel. The set of integers $\{1,2,\ldots,n\}$ is written as $I_n$, and the cardinality of a set $\mathcal{A}$ is written as $|\mathcal{A}|$. Denote the $N$ files as $W_1,W_2,\ldots,W_N$. Fix an integer parameter $t\in\{1,2,\ldots,K\}$ in the proposed scheme, then each file in our scheme is partitioned into ${K \choose t}$ segments of equal size. Each segment $W_{n,\mathcal{S}}$, where $n\in I_N$ and $\mathcal{S}\subseteq I_K$ with $|\mathcal{S}|=t$, is  assumed to be a symbol in $\mathbb{F}_{q^m}$ for some $q$ and $m$ sufficiently large. The parameters of $q$ and $m$ will be specified later. We reserve the calligraphic letter $\mathcal{S}$ for the purpose of enumerating some of the subsets of $I_K$ of cardinality $t$, without explicitly writing these conditions for notational simplicity. 

To present the general scheme, a few additional coding components are required. We first need a set of generic systematic linear MDS codes whose generator matrix has entries in $\mathbb{F}_q$ with parameters $(n_c,k_c)$, for all $n_c\geq k_c\geq 1$ and $n_c\leq q$; such codes can be found for any sufficiently large $q$, for example, using Cauchy matrix. We also allow the information symbols and coded symbols to be in $\mathbb{F}_{q^m}$, by taking the natural $\mathbb{F}_{q^m}$ finite field operation; this essentially boils down to writing the symbols as vectors length-$m$ in $\mathbb{F}_{q}$. 
Furthermore, fix the parameter  
\begin{align}
P={K-1\choose t-1}N,
\end{align}
in the linearized polynomial and also fix 
\begin{align}
P_o=2{K-1\choose t-1}N-{K-2\choose t-1}(N-1)
\end{align}
values $\theta_i\in \mathbb{F}_{q^m}$, $i=1,2,\ldots,P_o$, which are linearly independent in $\mathbb{F}_{q}$. This polynomial can be used to construct a $(P_o,P)$ systematic rank metric code as discussed in Section \ref{subsec:pre}; we shall refer to this code as $\mathcal{C}(P_o,P)$. We are now ready to present the general caching strategy.

\subsection{Caching Strategy}

The caching strategy of the proposed can be described as follows.
For user $k$, collect the file segment symbols:
\begin{align*}
\{W_{n,\mathcal{S}},\text{ for all } n\in I_N, \text{ and all } \mathcal{S} \text{ such that } k\in\mathcal{S}\}
\end{align*}
and encode it using the systematic rank metric code $\mathcal{C}(P_o,P)$; the parity symbols are then placed in the cache of user $k$. 

\subsection{Transmission Strategy When All Files Are Requested}

Fix a parameter $t\in \{1,2,\ldots, K-1\}$, let us first consider the case when all the files are being requested; the cases $t=0$ or $t=K$ are omitted for which the scheme is trivial. For a given set of file requests from all the users, we define
\begin{align}
&I^{[n]}\triangleq \{k\in I_K: \text{user $k$ requests file } W_n\},\nonumber\\
&\qquad\qquad\qquad\qquad\qquad\quad n=1,2,\ldots,N,
\end{align}
and $m_n=|I^{[n]}|\geq 1$, $n=1,2,\ldots,N$. Furthermore, define the complementary set $\bar{I}^{[n]}\triangleq I_K\setminus I^{[n]}$.

For each file $W_n$, we classify its segments $W_{n,\mathcal{S}}$ by its intersection with $\bar{I}^{[n]}$, and address them differently. More precisely, there are three steps of transmissions:
\begin{itemize} 
\item Step 1: All the file segments in the set $\{W_{n,\mathcal{S}}:\mathcal{S}\subseteq \bar{I}^{[n]}\}$ are transmitted uncoded directly; 
\item Step 2: For each subset $\mathcal{A}\subseteq \bar{I}^{[n]}$, where $|\mathcal{A}|=\max(1,t-m_n),\ldots,\min(t-1,K-m_n)$, we encode the set of file segments
\begin{align}
\mathcal{W}_{n,\mathcal{A}}\triangleq\{W_{n,\mathcal{S}}:\mathcal{S}\cap\bar{I}^{[n]}=\mathcal{A}\}
\end{align}
using a 
\begin{align}
\left(2{m_n\choose t-|\mathcal{A}|}-{m_n-1\choose t-|\mathcal{A}|-1},{m_n\choose t-|\mathcal{A}|}\right)=\left({m_n\choose t-|\mathcal{A}|}+{m_n-1\choose t-|\mathcal{A}|},{m_n\choose t-|\mathcal{A}|}\right)
\end{align} systematic MDS code (whose coding coefficients are in $\mathcal{F}_q$), and then transmit all the parity symbols; here we take the convention of ${n \choose k}=1$ when $k=0$.
\item Step 3: Encode all the file segments in the set $\mathcal{W}_{n,\emptyset}\triangleq\{W_{n,\mathcal{S}}:\mathcal{S}\subseteq {I}^{[n]}\}$ using a 
\begin{align}
\left(2{m_n\choose t}-{m_n-1\choose t-1},{m_n\choose t}\right)=\left({m_n\choose t}+{m_n-1\choose t},{m_n\choose t}\right)
\end{align}
 systematic MDS code  (whose coding coefficients are in $\mathcal{F}_q$), and then transmit all the parity symbols.
\end{itemize}

In fact we can even merge all the three steps by taking certain convention on degenerate MDS codes, however we keep them separate to facilitate understanding and analysis in the next section. For the required MDS codes to exist, a trivially sufficient finite field size is $q\geq 2{K-N+1 \choose \max(\lfloor (K-N+1)/2\rfloor,t)}$. For the required rank metric codes to exist, we can choose any $m\geq P_o$. 

It is clear that each file segment $W_{n,\mathcal{S}}$ either belongs to a singleton set $\{W_{n,\mathcal{S}}\}$ when $\mathcal{S}\subseteq \bar{I}^{[n]}$, or one of the sets $\mathcal{W}_{n,\mathcal{A}}$ for some $\mathcal{A}\subseteq\bar{I}^{[n]}$; in other words, for each $n$, the transmission strategy provides a partition of all the subset $\mathcal{S}$ for $S\subseteq I_K$ and $|\mathcal{S}|=t$ (and also induces a partition of all the file segments $W_{n,\mathcal{S}}$). For each $n$, we denote the mapping from a subset $\mathcal{S}$ to the corresponding subset that specifics the partition it belongs to as $\mathcal{A}_{I^{[n]}}(\mathcal{S})$, {\em i.e.,} $W_{n,\mathcal{S}}\in \mathcal{W}_{n,\mathcal{A}_{I^{[n]}}(\mathcal{S})}$.

\subsection{Transmission Strategy When Only Some Files Are Requested}

Again fix a parameter $t\in \{1,2,\ldots, K-1\}$, and consider the case when $N^*<N$ files are requested. Without loss of generality, let us assume that the first $N^*$ files are being requested, and  $I^{[n]}$, $m_n$ and $\bar{I}^{[n]}$ are defined similarly as in the last subsection, but only for $n=1,2,\ldots,N^*$. To describe the transmission strategy, we first find another set of \lq\lq{}enhanced demands\rq\rq{}, parametrized by $\dot{I}^{[1]},\dot{I}^{[2]},\ldots,\dot{I}^{[N]}$, where all files are being requested; {\em i.e.,} $|\dot{I}^{[n]}|\geq 1$ for $n=1,2,\ldots,N$. Additionally, these enhanced demands must satisfy the following properties:
\begin{itemize}
\item $|\dot{I}^{[n]}|=1$ for $n=N^*+1,\ldots,N$;
\item For any $k\in \{1,2,\ldots,K\}$, if $k\in I^{[n]}$, then either $k\in \dot{I}^{[n]}$, or $k\in \dot{I}^{[n\rq{}]}$, for some $n\rq{}\in \{N^*+1,\ldots,N\}$; for the latter case, denote the mapping from $n\rq{}$ to $n$ as $f(n\rq{})=n$, and denote the mapping from $n\rq{}$ to $k$ as $u(n\rq{})$.
\end{itemize}
We also write $|\dot{I}^{[n]}|=\dot{m}_n$ for simplicity.
The enhancement replaces some users\rq{} requests with requests for files that originally are not being requested, and each of these files is now being requested by only one user in the enhanced version. Note that this enhancement can always be found under the condition $N\leq K$. 

A set of counters need to be initialized before presenting the transmission strategy, which is given as
\begin{align}
\tau_{n,\mathcal{A}}\triangleq {\dot{m}_n-1\choose t-|\mathcal{A}|-1}, \qquad n=1,2,\ldots,N^* \mbox{  and   } \mathcal{A}\subseteq \bar{\dot{I}}^{[n]}.
\end{align}
Note that the set $\mathcal{A}$ can be $\emptyset$, and in fact in the proposed scheme we only need to consider the sets $\mathcal{A}$ where $|\mathcal{A}|\leq t-1$, though the definition is still valid for other cases, by taking the convention ${n \choose k}=0$ if $k<0$.

The transmission strategy is as follows:
\begin{itemize} 
\item For each file $W_n$, $n=1,2,\ldots,N^*$, transmit as described Step 1-3 for the \textbf{enhanced demands};
\item Step 4: for each $n$, $n=N^*+1,\ldots,N$, perform the following operations. For each $\mathcal{S}$, where $u(n)\notin \mathcal{S}$, reduce the counter $\tau_{f(n),\mathcal{A}_{\dot{I}^{[f(n)]}}(\mathcal{S})}$ by 1, and then transmit 
\begin{align}
\left\{\begin{array}{ll}
W_{f(n),\mathcal{S}}, & \mbox{if   } \tau_{f(n),\mathcal{A}_{\dot{I}^{[f(n)]}}(\mathcal{S})}\geq 0\\
W_{n,\mathcal{S}}, & \mbox{otherwise}
\end{array}\right..
\end{align}
\end{itemize}

\subsection{Revisiting the $(N,K)=(2,4)$ Example}

Let us revisit the example code for the $(2,4)$ case within the context of the general caching scheme. The two indexing methods now have the following mapping
\begin{align*}
A_1\rightarrow W_{1,\{1,2\}},\, A_2\rightarrow W_{1,\{1,3\}},\,A_3\rightarrow W_{1,\{1,4\}},\\
A_4\rightarrow W_{1, \{2,3\}},\, A_5\rightarrow W_{1,\{2,4\}},\,A_6\rightarrow W_{1,\{3,4\}},
\end{align*}
and similarly for file segments of file $B$.

The scheme presented earlier is for $t=2$. Though we did not utilize rank metric codes for this example, we can still derive the parameters $P_0=10$ and $P=6$, and thus $P_0-P=4$ symbols are generated and cached at each user. 

Now consider requests $(A,A,A,B)$, for which $m_1=3$  and $m_2=1$. It is clear that the uncoded transmission in the general scheme matches exactly what we have presented. Next consider the transmission in step 2 for $W_1=A$, $\mathcal{A}=\{4\}$ for which we have 
\begin{align}
\mathcal{W}_{1,\{4\}}=\{W_{1,\{1,4\}},W_{1,\{2,4\}},W_{1,\{3,4\}}\}=\{A_{3},A_{5},A_{6}\},
\end{align}
and the parities of a $(2{3 \choose 1}-{2 \choose 0},{3 \choose 1})=(5,3)$ MDS code are transmitted, which is exactly as that given previously, {\em i.e.}, the symbols $(A_3+2A_5+3A_6,A_3+3A_5+4A_6)$. In step 3, we have the following segments
\begin{align}
\mathcal{W}_{1,\emptyset}=\{W_{1,\{1,2\}},W_{1,\{1,3\}},W_{1,\{2,3\}}\}=\{A_{1},A_{2},A_{4}\},
\end{align}
and the parity symbol of a $(2{3 \choose 2}-{2 \choose 1},{3 \choose 2})=(4,3)$ MDS code is transmitted, which is exactly as that given previously, {\em i.e.}, the symbol $A_1+A_2+A_4$.
For file $W_2=B$, we can only take $|\mathcal{A}|=1$ in step 2 since $\max(1,t-m_2)=\min(t-1,4-1)=1$, however in this case, a $(2{1 \choose 1}-{0\choose 0},{1\choose 1})=(1,1)$ MDS code does not have any parity symbols, and thus no transmission of file $B$ is required in step 2; there is also no transmission of file $B$ in step 3. 

We can similarly walk through the example for $(N,K)=(3,4)$ using the general transmission strategy; this simple exercise is left to interested readers.

\section{Proof of the Main Theorem}
\label{sec:proof}
We establish the correctness and the performance of the caching scheme in three propositions, and Theorem \ref{theorem:main} follows directly from them. Two related issues are then  discussed, regarding the format of the cached linear combinations and the required field size of the code. Recall that we use $\mathcal{S}$ to enumerate file subsets $\mathcal{S}\subseteq I_K$ and $|\mathcal{S}|=t$.

\subsection{Correctness}

\begin{prop}
\label{prop:correctness}
For any $t\in\{1,2,\ldots,K-1\}$, the afore-given caching strategy can be used to satisfy any demands that request all files with the afore-given transmission strategy. 
\end{prop}
\begin{IEEEproof}

To show that any demands that request all $N$ files can be satisfied, we need consider any single user. Without loss of generality, we can consider the first user and assume it requests file $W_1$. Let us count the number of linear combinations he receives which consist of interference symbols in his cache in the first two transmission steps. 

In step 1, user 1 can collect all uncoded symbols for file $W_n$, $n=2,3,\ldots,N,$ in the form of 
\begin{align}
\{W_{n,\mathcal{S}}: 1\in\mathcal{S}\subseteq \bar{{I}}^{[n]}\},
\end{align}
and there are a total of
\begin{align}
\tilde{T}^{(1)}=\sum_{n=2}^N{K-{m}_{n}-1 \choose t-1}
\end{align}
such symbols, where we have taken the convention ${n \choose k}=0$ when $n<k$.

In step 2, user 1 collects linear combinations of $W_n$, $n=2,3,\ldots,N$, however only those in the following form. For each such $n$, and each subset $\mathcal{A}\subseteq \bar{I}^{[n]}$ such that $\max(1,t-m_n)\leq |\mathcal{A}|\leq \min(t-1,K-m_n)$ and moreover $1\in \mathcal{A}$, user 1 collects the parity symbols of encoding $\mathcal{W}_{n,\mathcal{A}}$ using the systematic MDS code. Thus user 1 collects a total of 
\begin{align}
\tilde{T}^{(2)}&=\sum_{n=2}^N\sum_{j=\max(1,t-m_{n})}^{\min(t-1,K-m_{n})}{K-m_{n}-1 \choose j-1}{m_{n}-1 \choose t-j}
\end{align}
such symbols. 

User 1 now has collected $\tilde{T}^{(1)}+\tilde{T}^{(2)}$ useful symbols, and has in his cache $P_o-P$ symbols of the same basis. Observe for the summands in $\tilde{T}^{(1)}$ and $\tilde{T}^{(2)}$, we have
\begin{align}
\sum_{j=\max(1,t-m_{n})}^{\min(t-1,K-m_{n})}{K-m_{n}-1 \choose j-1}{m_{n}-1 \choose t-j}
+{K-m_{n}-1 \choose t-1}={K-2 \choose t-1},
\end{align}
because the left hand side is simply all the possible ways of choosing $t-1$ balls in a total of $K-2$ balls, however counted when these balls are partitioned into two groups of size $K-2-(m_{n}-1)$ and $m_{n}-1$, respectively. It follows
\begin{align}
&\tilde{T}^{(1)}+\tilde{T}^{(2)}+P_o-P=P.
\end{align}
These $P$ linear combinations, which can be represented as the product of the length-$P_o$ output (both systematic and parity symbols) of the rank metric code $\mathcal{C}(P_o,P)$ and a matrix $G$ of size $P_o\times P$. Recall the systematic rank metric code we used to encode the $P$ file segments in user 1's cache, and by Lemma \ref{lemma:fullrank}, as long as the matrix $G$ is full rank, all the $P$ segments can be recovered. This fact is proved in the appendix, but an outline of the proof is given here. We recognize that if the columns and rows of the matrix $G$ are rearranged to 
\begin{itemize}
\item Group the file segments $W_{1,\mathcal{S}}$ in user 1's cache together;
\item For each $n=2,3,\ldots,N$, group the segments of $\{W_{n,\mathcal{S}}: 1\in \mathcal{S}\subseteq \bar{I}^{[n]}\}$ together;
\item For each $n=2,3,\ldots,N$, and for each subset $\mathcal{A}\subseteq \bar{I}^{[n]}$ such that $\max(1,t-m_n)\leq |\mathcal{A}|\leq \min(t-1,K-m_n)$ and moreover $1\in \mathcal{A}$, group the segments of $\mathcal{W}_{n,\mathcal{A}}$ together,
\end{itemize}
then the resulting matrix is block diagonal, and each block is either of size $1\times 1$ with entry $1$ or full rank because they are columns of generator matrices of MDS codes. Thus the matrix $G$ is indeed full rank. 

Thus user 1 can eliminate the interferences in its cached contents, and recover all the file segments of $W_{1,\mathcal{S}}$ that are already present in its cache. It remains to show that all the file segments $W_{1,\mathcal{S}}$ that are not present in his cache can also be recovered. 

First, observe that in step 1, user 1 can collect all uncoded $W_1$ file segments that are not in the cache of any users $k\in I^{[1]}$, {\em i.e.,} $\{W_{1,\mathcal{S}}:1\notin\mathcal{S}\subseteq \bar{I}^{[1]}\}$. As mentioned earlier, in step 2 after eliminating the interference, user 1 can recover all $W_{1,\mathcal{S}}$ for $\mathcal{S}$ such that $1\in\mathcal{S}$. Furthermore, for each subset $\mathcal{A}\subseteq \bar{I}^{[1]}$ such that $\max(1,t-m_1)\leq |\mathcal{A}|\leq \min(t-1,K-m_1)$, user 1 can collect the parity symbols of encoding $\mathcal{W}_{1,\mathcal{A}}$ using the $\left(2{m_1\choose t-|\mathcal{A}|}-{m_1-1\choose t-|\mathcal{A}|-1},{m_1\choose t-|\mathcal{A}|}\right)$ systematic MDS code. Since user 1 has in its cache ${m_1-1\choose t-|\mathcal{A}|-1}$ of the total ${m_1\choose t-|\mathcal{A}|}$ symbols of $\mathcal{W}_{1,\mathcal{A}}$, together with the collected parity symbols, he can recover all ${m_1\choose t-|\mathcal{A}|}$ symbols in this set. Thus after step 2, user 1 can also recover all file segments $W_{1,\mathcal{S}}$ where $\mathcal{S}$ has elements in both $I^{[1]}$ and $\bar{I}^{[1]}$. The only missing segments are some in the set $\{W_{1,\mathcal{S}}: 1\notin\mathcal{S}\subseteq I^{[1]}\}$. However, step 3 transmits the parities of a $\left(2{m_1\choose t}-{m_1-1\choose t-1},{m_1\choose t}\right)$ MDS code that encodes all $\{W_{1,\mathcal{S}}:\mathcal{S}\subseteq {I}^{[1]}\}$, and since user 1 already has ${m_1-1\choose t-1}$ elements, he can thus also recover the rest of the symbols in this set. At this point, we can conclude that user 1 can recover all file segments of $W_1$, which completes the proof. 
\end{IEEEproof}

\begin{prop}
\label{prop:correctnessenhanced}
For any $t\in\{1,2,\ldots,K-1\}$, the afore-given caching strategy can be used to satisfy any demands that request a strict subset of all the files with the afore-given transmission strategy. 
\end{prop}

The proof of this proposition can be intuitively explained as follows. When we replace a file demand $W_i$ in the enhanced demands with a demand $W_j$, the effect of the not transmitting the file segments involving $W_i$ in the first three steps needs to compensated.  In order to do so, let us examine the roles that these $W_i$ transmissions play: firstly, they are used to eliminate the interferences by $W_i$ at certain other users, and secondly, they are used to provide the missing segments to the single user that was requesting $W_i$ in the enhanced demands. Our strategy is to transmit the corresponding segments from $W_j$ instead of sending the segments from $W_i$. With such substituted transmissions, the first role can be fulfilled as long as it is not a redundant transmission, and we rely on the counter $\tau_{n,\mathcal{A}}$ to avoid any such redundancy. The second role can clearly also be fulfilled by any such non redundant transmissions. When a transmission of the file segment from $W_j$ is indeed redundant, we can safely conclude that the second role has already been fulfilled by previous transmissions, and thus transmitting this segment of $W_i$ is now sufficient to serve the first role alone.  The proof below makes this intuition more rigorous.

\begin{IEEEproof}
Without loss of generality, we only need to consider the first user and assume his request is for file $W_1$. Two cases need to be examined: the first case is when in the enhanced demands, the first user was also requesting file $W_1$; the second case is when in the enhanced demands, the first user was requesting $n^*$, {\em i.e., } $f(n^*)=1$ and $u(n^*)=1$, for some $n^*\in\{N^*+1,\ldots,N\}$. 

Let us consider the proof for the first case, which is similar to the proof for the Proposition \ref{prop:correctness}. In step 1, user 1 collects all uncoded symbols for file $W_n$, $n=2,3,\ldots,N^*,$ in the form of 
\begin{align}
\{W_{n,\mathcal{S}}: 1\in\mathcal{S}\subseteq \bar{\dot{I}}^{[n]}\},
\end{align}
and there are a total of
\begin{align}
\tilde{\dot{T}}^{(1)}=\sum_{n=2}^{N^*}{K-\dot{m}_{n}-1 \choose t-1}
\end{align}
such symbols.

In step 2, user 1 collects linear combinations of $W_n$, $n=2,3,\ldots,N^*$, however only those in the following form. For each such $n$, and each subset $\mathcal{A}\subseteq \bar{\dot{I}}^{[n]}$ such that $\max(1,t-\dot{m}_n)\leq |\mathcal{A}|\leq \min(t-1,K-\dot{m}_n)$ and moreover $1\in \mathcal{A}$, user 1 collects the parity symbols of encoding $\mathcal{W}_{n,\mathcal{A}}$ using the systematic MDS code. Thus user 1 collects a total of 
\begin{align}
\tilde{\dot{T}}^{(2)}&=\sum_{n=2}^{N^*}\sum_{j=\max(1,t-\dot{m}_{n})}^{\min(t-1,K-\dot{m}_{n})}{K-\dot{m}_{n}-1 \choose j-1}{\dot{m}_{n}-1 \choose t-j}
\end{align}
such symbols. 

In step 4, user 1 collects for each $n=N^*+1,\ldots,N$, for any $\mathcal{A}\subseteq \bar{\dot{I}}^{[n]}$ where $|\mathcal{A}|=t-1$ and $1\in\mathcal{A}$, either $W_{n,\mathcal{A}\cup\{u(n)\}}$ or $W_{f(n),\mathcal{A}\cup\{u(n)\}}$, whichever was transmitted in step 4. Note that in this case $u(n)\neq 1$ for any $n=N^*+1,\ldots,N$, which implies that $|\mathcal{A}\cup\{u(n)\}|=t$. Thus user 1 collects another total of 
\begin{align}
\tilde{\dot{T}}^{(4)}&=(N-N^*){K-2 \choose t-2}
\end{align}
uncoded symbols. 

User 1 now has collected $\tilde{\dot{T}}^{(1)}+\tilde{\dot{T}}^{(2)}+\tilde{\dot{T}}^{(4)}$ useful symbols, and has in his cache $P_o-P$ symbols of the same basis. It is seen that
\begin{align}
&\tilde{T}^{(1)}+\tilde{T}^{(2)}+\tilde{T}^{(4)}+P_o-P=P.
\end{align}
These $P$ linear combinations, which can again be represented as the product of the length-$P_o$ output (both systematic and parity symbols) of the rank metric code $\mathcal{C}(P_o,P)$ and a matrix $G^*$ of size $P_o\times P$. As long as the matrix $G^*$ is full rank, user 1 can recover all the file segments $W_{1,\mathcal{S}}$ where $1\in \mathcal{S}$, and the rest of file segments from $W_1$ can be recovered as in the case of Proposition \ref{prop:correctness}. The fact of the matrix $G^*$ being full rank is obvious for the similar reason that the $G$ matrix is full rank under the enhanced demands; in fact, since the transmissions in step 4 are all uncoded, the full-rank property directly follows from the full-rank property of the corresponding matrix with the enhanced demands. 
 
Now let us now  consider the second case, where user 1 is demanding $W_1$, but in the enhanced demands, he was requesting file $n^*$ for some $n^*\in \{N^*+1,\ldots,N\}$. By a similar argument as above, user 1 can recover all segments $W_{1,\mathcal{S}}$ present in his cache, {\em i.e.,} for $W_{1,\mathcal{S}}$ where $1\in\mathcal{S}$, by eliminating the interferences. More precisely, in step 1, user 1 collects all uncoded symbols for file $W_n$, $1,2,\ldots,N^*,$ in the form of 
\begin{align}
\{W_{n,\mathcal{S}}: 1\in\mathcal{S}\subseteq \bar{\dot{I}}^{[n]}\},
\end{align}
and there are a total of
\begin{align}
\tilde{\dot{T}}^{(1\rq{})}=\sum_{n=1}^{N^*}{K-\dot{m}_{n}-1 \choose t-1}
\end{align}
such symbols.
In step 2, user 1 collects linear combinations of $W_n$, $n=1,2,\ldots,N^*$, however only those in the following form. For each such $n$, and each subset $\mathcal{A}\subseteq \bar{\dot{I}}^{[n]}$ such that $\max(1,t-\dot{m}_n)\leq |\mathcal{A}|\leq \min(t-1,K-\dot{m}_n)$ and moreover $1\in \mathcal{A}$, user 1 collects the parity symbols of encoding $\mathcal{W}_{n,\mathcal{A}}$ using the systematic MDS code. Thus user 1 collects a total of 
\begin{align}
\tilde{\dot{T}}^{(2\rq{})}&=\sum_{n=1}^{N^*}\sum_{j=\max(1,t-\dot{m}_{n})}^{\min(t-1,K-\dot{m}_{n})}{K-\dot{m}_{n}-1 \choose j-1}{\dot{m}_{n}-1 \choose t-j}
\end{align}
such symbols. In step 4, user 1 collects for each $n=N^*+1,\ldots,n^*-1,n^*+1,\ldots,N$, for any $\mathcal{A}\subseteq \bar{\dot{I}}^{[n]}$ where $|\mathcal{A}|=t-1$ and $1\in\mathcal{A}$, either $W_{n,\mathcal{A}\cup\{u(n)\}}$ or $W_{f(n),\mathcal{A}\cup\{u(n)\}}$, whichever was transmitted in step 4. Thus user 1 collects another total of 
\begin{align}
\tilde{\dot{T}}^{(4\rq{})}&=(N-N^*-1){K-2 \choose t-2}
\end{align}
uncoded symbols. User 1 now has collected $\tilde{\dot{T}}^{(1)}+\tilde{\dot{T}}^{(2)}+\tilde{\dot{T}}^{(4)}$ useful symbols, and has in his cache $P_o-P$ symbols of the same basis. It is seen that
\begin{align}
&\tilde{T}^{(1\rq{})}+\tilde{T}^{(2\rq{})}+\tilde{T}^{(4\rq{})}+P_o-P=P.
\end{align}

It only remains to show that for the second case, the transmissions in Step 4 suffice to provide any missing segments of $W_1$ in user 1's cache, possibly jointly with transmissions from $W_1$ in the first three steps. This is rather straightforward, since all file segments $W_{1,\mathcal{S}}$'s with $1\not\in \mathcal{S}$ are transmitted uncoded in Step 4, unless $\tau_{1,\mathcal{A}_{\dot{I}^{[1]}}(\mathcal{S})}<0$; when the latter scenario occurs, a total of 
${\dot{m}_1-1\choose t-|\mathcal{A}_{\dot{I}^{[1]}}(\mathcal{S})|-1}$ 
uncoded symbols have already been transmitted in the set $\mathcal{W}_{1,\mathcal{A}_{\dot{I}^{[1]}}(\mathcal{S})}$, and together with the
${\dot{m}_n-1\choose t-|\mathcal{A}_{\dot{I}^{[1]}}(\mathcal{S})|}$ parity symbols encoding the set $\mathcal{W}_{1,\mathcal{A}_{\dot{I}^{[1]}}(\mathcal{S})}$ which were transmitted in the first three steps, user 1 can indeed recover all ${\dot{m}_n\choose t-|\mathcal{A}_{\dot{I}^{[1]}}(\mathcal{S})|}$ symbols in $\mathcal{W}_{1,\mathcal{A}_{\dot{I}^{[1]}}(\mathcal{S})}$. Thus user 1 is able to recover all segments of $W_{1}$, and the proof is complete. 
\end{IEEEproof}

\subsection{Performance}

\begin{prop}
\label{prop:degenerate}
For any $t\in\{1,2,\ldots,k-1\}$, the afore-given caching strategy and transmission strategy achieve the memory-transmission pair
\begin{align}
(M,R)=\left(\frac{t[(N-1)t+K-N]}{K(K-1)},\frac{N(K-t)}{K}\right).
\end{align}
\end{prop}
\begin{IEEEproof}
Recall each file of unit size is partition into ${K \choose t}$ segment symbols, and each user caches $P_o-P$ symbols, and thus the memory usage is straightforwardly to calculate. It remains to calculate the total number of transmitted symbols. 

We only need to consider the first three steps of transmission when all files are being requested, since for the other cases where only a subset of files are requested, each transmission in step 4 corresponds to exactly one transmission in step 3 for the enhanced demands, and thus the rate remains the same as for the case of the enhanced demands. 

Clearly, in step 1, the total number of transmitted uncoded symbols of file $W_n$ is
\begin{align}
T^{(1)}_n={K-m_n \choose t}.
\end{align}
In step 2, the total number of transmitted linear combinations of file $W_n$ is given as
\begin{align*}
T^{(2)}_n=\sum_{j=\max(1,t-m_n)}^{\min(t-1,K-m_n)} {K-m_n \choose j} {m_{n}-1 \choose t-j}.
\end{align*}
In step 3, the total number of transmitted linear combinations of file $W_n$ is given as
\begin{align*}
T^{(3)}_n={m_n-1 \choose t}.
\end{align*}
Note that
\begin{align*}
T^{(1)}_n+T^{(2)}_n+T^{(3)}_n={K-1\choose t},
\end{align*}
because it is all the ways of choosing $t$ balls in a total of $K-1$ balls. Thus the total transmissions amount to $N{K-1\choose t}$ symbols. The proof can now be completed with a simple normalization by the number of segments in each file. 
\end{IEEEproof}

\subsection{The Semi-Systematic Variant of the Caching Strategy}

The general caching strategy we provide does not enforce any special structure on the linear combinations, unlike the code given in the $(2,4)$ example. However, even for general parameters $(N,K)$ and the same range of parameter $t$, we can indeed choose to use the semi-systematic format. More precisely, the first ${K-1\choose t-1}$ semi-systematic symbols in the cache of user $k$ are
\begin{align}
\sum_{n=1}^NW_{n,\mathcal{S}},\quad k\in \mathcal{S},
\end{align}
where the addition is in finite field $\mathbb{F}_{q^m}$. 
Moreover, we use the same parameter $P$, but choose
\begin{align}
P'_o={K-1\choose t-1}(2N-1)-{K-2\choose t-1}(N-1),
\end{align}
and construct a $(P'_o,P)$ systematic rank metric code, which is denoted as $\mathcal{C}(P'_o,P)$. The local parity symbols stored in user $k$'s cache are the parity symbols when encoding the set of file segment symbols $\{W_{n,\mathcal{S}}:n=1,2,\ldots,N,k\in \mathcal{S}\}$ using $\mathcal{C}(P'_o,P)$. The transmission strategy remains the same. 

In order to prove the correctness of this caching variant, we only need to show that the corresponding matrix $G'$, similarly as in the proof of Proposition \ref{prop:correctness}, is also full rank. This is again rather immediate. Since the only difference is the columns corresponding to the semi-systematic symbols in the cache. However, it is easily seen that although the matrix $G'$ is no longer block diagonal after the rearrangement of columns and rows, the new columns has non-zero entries on rows corresponding to $W_{1,\mathcal{S}}$ (in fact it has an identity matrix if we restrict it to these columns and rows with proper row and column indexing), while no other columns in $G'$ have non-zero entries on these rows. Thus indeed this variant of caching strategy is also valid; a more precise proof is given in the appendix. 

We choose to present the general construction in the last section instead of this variant directly in order to emphasize the fact that the semi-systematic format is not fundamentally important in our construction. Note that in the semi-systematic variant, the bound on the parameter $m$ can be made smaller, since the parameters of the rank metric code are reduced: choosing $m\geq P'_o$ suffices here.

\subsection{Reducing the Field Size with Generic Linear Codes}

In the proposed code construction, we rely on rank metric codes to guarantee certain full rank properties, and the overall code design problem essentially reduces to a rank counting problem on the proper basis. However, one obvious disadvantage of using rank metric codes in the construction is that the size of the field $\mathbf{F}_{q^m}$ needs to be quite large. We can in fact replace the rank metric code with a generic systematic linear code, and directly require the full rank properties to hold. In this section, we provide such a simple argument and show that a reduced field size is sufficient. 

Let us consider the cache encoding for the $k$-th user. A total of $P$ symbols are present at this user, and a total of $P_o-P$ parity symbols are generated during the encoding. In this subsection, we shall assume that the entries of this $P\times(P_o-P)$ encoding matrix are from $\mathbb{F}_q$, {\em i.e.,} the same finite field as the set of MDS codes. Denote this matrix as $G_k$, and its entry on the $i$-th row and $j$-th column as $g_{k,i,j}$, which are to be determined; note that this code is not necessarily a rank-metric code any longer. 

Consider a specific set of demands $(d_1,d_2,\ldots,d_K)$, ({\em i.e.}, the $k$-th user demands file $d_k$), where all files are requested. In the delivery phase, the symbols user-$k$ collects during Step 1 and Step 2 are linear combinations of all the symbols present at this user. This can be represented also by a $P\times (\tilde{T}^{(1)}+\tilde{T}^{(2)})$ encoding matrix $G'_{k,(d_1,d_2,\ldots,d_K)}$. The full rank condition in the proof of Proposition \ref{prop:correctness} essentially requires that the $P\times P$ matrix $[G_k, G'_{k,(d_1,d_2,\ldots,d_K)}]$ being full rank. The determinant of the matrix $[G_k, G'_{k,(d_1,d_2,\ldots,d_K)}]$ can be expressed as a function of the coefficients $g_{k,i,j}$'s, {\em i.e.}
\begin{align*}
\det(G_n, G'_{n,(d_1,d_2,\ldots,d_K)})=f_{n,(d_1,d_2,\ldots,d_K)}(\{g_{n,i,j}\}).
\end{align*}

By the proof of Proposition \ref{prop:degenerate}, the full rank condition for demands where only a subset of the files are requested is implied by the full rank condition for the enhanced demands. Thus as long as the following polynomial has a non-zero solution, then the choice of coefficients $\{g_{k,i,j}\}$ is valid
\begin{align}
\prod_{k=1}^K \prod_{\substack{(d_1,d_2,\ldots,d_K):\\\text{all files requested}}}f_{k,(d_1,d_2,\ldots,d_K)}(\{g_{k,i,j}\}).\label{eqn:largepoly}
\end{align}

We can now invoke the following lemma. 
\begin{lemma} \cite{Alo} \label{lem:nullstellansatz} ({\em Combinatorial Nullstellansatz}) Let $\mathbb{F}$ be a field, and let $f = f(x_1 , \cdots , x_n)$ be a polynomial in $\mathbb{F}[x_1, \cdots , x_n ]$. Suppose the degree $\text{deg}(f)$ of $f$ is expressible in the form $\sum_{i=1}^{n} t_i$, where each $t_i$ is a non-negative integer and suppose that the coefficient of the monomial term $\prod_{i=1}^{n} x_i^{t_i}$  is nonzero. Then if $S_1 ,\ldots , S_n$ are  subsets of $\mathbb{F}$ with sizes $|S_i|$ satisfying $|S_i| > t_i$, then there exist elements $s_1 \in S_1, s_2 \in S_2 \ldots , s_n \in S_n$ such that $f(s_1, s_2 , \cdots , s_n ) \neq 0$.
\end{lemma}

In this lemma above, the condition that the coefficient of the monomial term $\prod_{i=1}^{n} x_i^{t_i}$  is nonzero is equivalent to requiring $f = f(x_1 , \cdots , x_n)$ to be not identically zero. We note that $f_{n,(d_1,d_2,\ldots,d_K)}(\{g_{n,i,j}\})$ is indeed not identically zero, because the code construction previously given provides a non-zero assignment. 

Since the degree of any indeterminate in each of $f_{k,(d_1,d_2,\ldots,d_K)}(\{g_{n,i,j}\})$ is $1$, the maximum among the degrees of a single indeterminate of the polynomial (\ref{eqn:largepoly}) is upper bounded by the total number of demands where all files are requested, which is given by $S(K,N)N!$. Here
\begin{align}
S(K,N)=\frac{1}{N!}\sum_{j=1}^N(-1)^{N-j}{N \choose j}j^K,
\end{align}
is the Sterling number of the second kind  \cite{Knuth:book}, which counts the number of ways to partition a set of $K$ objects into $N$ non-empty subsets. 
Hence by Lemma~\ref{lem:nullstellansatz}, it is possible to find a suitable assignment for $\{ g_{n,i,j} \}$, if the entries are picked from a finite field $\mathbb{F}_q$ with $q> S(K,N)N!$. Alternatively, we can simply count the total number of demands, instead those where all files are requested, and this leads to a looser bound $N^K$ on the field size. 

We suspect that this bound can be further reduced through a more careful analysis of the matrix structure, though so far our effort toward this goal does not bear much fruit. Moreover, by allowing a larger number of cached symbols per user, codes in even smaller finite fields may be possible.

\section{Conclusion}
\label{sec:conclusion}

We proposed a new coding scheme for the caching problem when $N\leq K$, based on a combination of rank metric codes and MDS codes. The performance of the scheme has a particularly simple form, and it provides new tradeoff points beyond known what are known in the literature. Compared to known coded caching schemes, the proposed scheme uses coding for both caching and delivery, as well as larger finite field instead of finite field of cardinality 2.

An immediate variation of the proposed scheme is its decentralized counterpart, motivated by the investigation of the decentralized caching scheme \cite{MaddahAliNiesen:14Networking}, which is a variation of the centralized caching scheme in \cite{MaddahAliNiesen:14}.  We suspect that our scheme can also be extended to decentralized scenarios where certain random linear combinations of the file segments are cached, however it remains to be seen whether the performance such attained is still competitive.

\section*{Appendix: Full Rank of Matrix $G$ and $G'$}

The key to the proof is to express the matrix $G$ of size $P_o\times P$ in a more structured manner. For this purpose, let us again consider the cache and decoding process at user 1. First rearrange the systematic and parity symbols of the code $\mathcal{C}(P_o,P)$, such that the $P_o-P$ cached symbol are indexed in the set $I_{P_o-P}$; similarly we arrange the columns the $G$ such that its first $P_o-P$ columns correspond to these cached symbols. The next rows and columns correspond to the symbols that user 1 collected during the step 1 transmission
\begin{align*}
\{W_{n,\mathcal{S}}: 1\in\mathcal{S}\subseteq \bar{I}^{[n]}\},\quad n=2,3,\ldots,N, 
\end{align*}
and there are a total of $\sum_{n=2}^N\tilde{T}^{(1)}_n$ such symbols. 

The next rows and columns correspond to a fixed $n\in \{2,3,\ldots,N\}$ and a fixed subset $\mathcal{A}\subseteq \bar{I}^{[n]}$ where $\max(1,t-m_n)\leq |\mathcal{A}|\leq \min(t-1,K-m_n)$ and moreover $1\in \mathcal{A}$. Denote the parity check portion of generator matrix of the $\left(2{m_n\choose t-|\mathcal{A}|}-{m_n-1\choose t-|\mathcal{A}|-1},{m_n\choose t-|\mathcal{A}|}\right)$ systematic MDS as $Q_{n,\mathcal{A}}$, which has dimension ${m_n\choose t-|\mathcal{A}|}\times\left[{m_n\choose t-|\mathcal{A}|}-{m_n-1\choose t-|\mathcal{A}|-1}\right]$, and it is full rank since it is part of a generator matrix of an MDS code and it has less columns than rows. 

Now the matrix $G$ can be written in the following form
\begin{align}
G=\begin{bmatrix} I &                                    &                                    &       &                                    \\  
                            &  Q_{2,\mathcal{A}_{2,1}} &                                   &       &                                    \\
                            &                                     &Q_{2,\mathcal{A}_{2,2}} &       &                                    \\
                            &                                     &                                   & ...   &                                    \\
                            &                                     &                                   &       &Q_{N,\mathcal{A}_{N,L_N}}
\end{bmatrix}
\end{align}
where the identity matrix at the top-left has dimension $(P_o-P+\sum_{n=2}^N\tilde{T}^{(1)}_n)\times(P_o-P+\sum_{n=2}^N\tilde{T}^{(1)}_n)$, and we have enumerated the aforementioned matrix $\mathcal{A}$'s for each $n$ by using the subscript as $\mathcal{A}_{n,\ell}$, and $L_N$ is the total number of such subsets $\mathcal{A}$ when $n=N$. It is now clear that the matrix $G$ is block diagonal and each block is full rank, and thus $G$ indeed has full rank.

For the semi-systematic variant of the caching scheme, the matrix $G'$ is slightly different. First index the symbols
\begin{align} 
\{W_{1,\mathcal{S}}:1\in\mathcal{S}\}
\end{align}
using the set $I_{{K-1\choose t-1}}$, and rearrange the columns and rows of $G'$ such that they correspond to the top ${K-1\choose t-1}\times {K-1\choose t-1}$ submatrix using the same order. Next rearrange the systematic and parity symbols of the code $\mathcal{C}(P'_o,P)$, such that the $P'_o-P$ cached symbol correspond to the next $P'_o-P$ columns and rows. The rest of the $G'$ matrix is arranged exactly as for the case $G$. It is now clear that the matrix $G'$ has the following form
\begin{align}
G'=\begin{bmatrix} 
I_a                                 &    &                                     &                                  &       &                                    \\
                                     &I_b&                                     &                                  &       &                                    \\  
F_{2,\mathcal{A}_{2,1}}    &    &  Q_{2,\mathcal{A}_{2,1}} &                                   &       &                                    \\
F_{2,\mathcal{A}_{2,2}}    &    &                                     &Q_{2,\mathcal{A}_{2,2}} &       &                                    \\
                                     &    &                                     &                                   & ...   &                                    \\
F_{N,\mathcal{A}_{N,L_N}}&      &                                     &                                   &       &Q_{N,\mathcal{A}_{N,L_N}}
\end{bmatrix}
\end{align}
where the identity matrix $I_a$ is of dimension ${K-1\choose t-1}\times {K-1\choose t-1}$, and  the identity matrix $I_b$ is of dimension $(P'_o-P+\sum_{n=2}^N\tilde{T}^{(1)}_n)\times(P'_o-P+\sum_{n=2}^N\tilde{T}^{(1)}_n)$, and the $F_{n,\mathcal{A}_{n,\ell}}$ matrices have some nonzero entries but their exact forms are not important here; the other off block-diagonal entries are all zeros. It is now clear that the matrix $G'$ also has full rank. \qed

\bibliographystyle{IEEEtran}

\end{document}